Coupled structural and magnetic properties of ferric fluoride nanostructures
part I: a Metropolis atomistic study

B. Fongang, I. Labaye, F. Calvayrac, and J.M. Greneche
*Laboratoire de Physique de l'Etat Condensé, CNRS UMR 6087, IRIM2F CNRS FR 2575 Université du Maine, 72085 Le Mans, Cedex 9, France*
S. Zékeng
*Laboratoire de Sciences des Matériaux, Département de Physique, Université de Yaoundé I, BP 812 Yaoundé, Cameroun*

Abstract: A modified Metropolis atomistic simulation is proposed to model the structure of grain boundaries (GBs) and interfaces in ionic nanostructured systems and is applied to the magnetically interesting case of iron trifluoride ($FeF_3$). We chose long-range interatomic potentials adjusted on experimental results, and adapted a previously established Monte Carlo scheme consisting of various modifications of the simulated annealing/ Metropolis algorithm. Atomic structures of twisted and tilted GBs as a function of the relative disorientation of the grains have been achieved yielding close to experimentally measured properties. This approach takes into account the structure of the grains far from the interface in order to constrain the relative orientation of the grains, without any periodic boundary conditions. One concludes that a long-range coulombic falloff of the interatomic potentials is necessary to obtain GB structures presenting a correct local topology but with a smooth transition from crystalline to amorphous states. The structural features are finally discussed in terms of topological aspects and local magnetic structure.

1. Introduction

Different solid-state processing techniques based on chemical and physical routes, such as spray forming, *in situ* consolidation of aggregates, high energy ball milling allow the so-called nanostructured materials to be prepared. High-energy ball milling, which involves successive fragmentation and welding stages of a microcrystalline powder into a dense random assembly of nanocrystalline grains, gives rise to tunable and reproducible nanostructured powders [1-4]. This non-equilibrium state is obviously not the lowest energy one as it would be a quasi-infinite perfectly crystalline bulk at a close to zero temperature and pressure. In those latter conditions, however, the nanostructured state is metastable on macroscopic time scales [5, 6] and can be experimentally studied, but originates a modeling problem due to this metastability, which remains hard to address with a brute force simulation such as direct molecular dynamics which would, on the one hand, be unable to address such large time scales, and on the other hand leads to a perfect bulk or cluster if done without constraints or modifications. These nanostructured systems contain disordered interfaces or grain boundaries (GBs) on a nanometer scale (up to a few nm), when the grains themselves can be almost crystalline on a much larger scale [7-10]. The atomic fraction localized in those GBs, which is dependent on the conditions of synthesis, increases when either their thickness increases or/and the grain size decreases. Furthermore, rough interfaces, which are also observed in multilayers, prepared by physical processes such as molecular-beam epitaxy, pulsed laser ablation, behave as GBs. It is important to emphasize that the GBs significantly influence some physical properties in nanostructured materials: for example thermal, and electrical conductivity, coercive field and magnetic losses, and in multilayers transport properties in spin electronics and superconductive devices. Indeed, the switching characteristics are expected to be dependent on the electronic properties of the interfaces, i.e. the structural matching, in the case of resistive switching devices based on transition metal oxides [11, 12]. Consequently, the structural modeling of the GBs becomes a relevant challenging task [13-16].

When their proportion exceeds about 15% in volume, the presence of GB can be evidenced by means of diffraction experiments, spectroscopic techniques and magnetic measurements, and they can be observed by high-resolution transmission electron microscopy. In previous studies [4, 17], [57]Fe Mössbauer spectrometry has clearly evidenced the existence of two components of comparable volume fraction in the case of nanostructured fluoride powders. Indeed these ionic nanostructures consist of two components with different magnetic behaviors [4, 17]: the rhombohedral ferric fluoride is transformed after milling into a structure resulting from a pseudo-cubic packing and a random packing of corner sharing $FeF_6$ octahedral units assigned to antiferromagnetic crystalline grains and speromagnetic-like GBs, respectively. It is important to remember that both the cationic network involving odd and even-membered rings and the presence of antiferromagnetic superexchange interactions gives rise to non-collinear magnetic structures. For those reasons, the polymorphism of crystalline and amorphous phases of ferric fluorides nicely illustrates this purpose of atomistically describing a real ionic nanostructured system with interesting magnetic properties. $FeF_3$ presents several phases [5], the rhombohedral one with r-3c symmetry being the more stable at ambient temperature and presenting at 407 °C a reversible phase transition to a cubic phase analogous to rhenium oxide ReO3. In addition, $FeF_3$ exhibits a hexagonal (HTB) phase with space group P6/mmm and a pyrochlore-like phase with space group Fd3m, both containing 3-membered iron rings originating non collinear magnetic structures involving 3 and 4 antiferromagnetic sub lattices, respectively. Finally, the speromagnetic amorphous forms have to be described by means of random packing of corner sharing $FeF_6$ octahedral units [18-20].

In the present study we propose a numeric methodology allowing the structural modeling of grain boundaries in such an ionic system. We restrict ourselves to the following system: a double boundary constrained in between grains which consist of pure rhombohedral ferric fluoride of a 10 nanometer size on average, as experimentally measured in powders prepared by high energy ball grinding [17]. Eventual dislocations or stresses inside the grains are supposed to have relaxed (or the grains to have been slightly annealed) and assuming they have no impact on the structure of the GB, are then neglected. Thus, we have to address the atomic modeling of such a system in an out of equilibrium metastable state, induced by the mismatch of the individual crystalline structure of the grains far from the interfaces.

2. Theoretical Aspects

Ionic systems such as ferric fluoride can be very well described by the use of two-body interatomic potentials with a long-range coulombic form without cutoff and a short-range corrective term with various degrees of refinement from the Born-Mayer equation to the more elaborated approaches used in GULP [21] for instance. The cut-off of the long-range coulombic part is crucial to avoid prohibitive computational cost even if there are several numerical techniques (N-tree, Fast Multipole, etc) to reduce this cost [21-26]. However it turns out that if the cut-off is smaller than the width of two octahedral units the various crystalline forms of the system are stable under a moderate annealing. But starting from a random configuration of the atoms it is not possible to reproduce long-range order beyond one octahedral unit and one ends up with an amorphous state as in [26]. This state can be experimentally obtained but it is not possible to describe the disordered structure of GB with long-range order using such potentials. Therefore, for the Fe-F and F-F interactions (Fe atoms are too far from each other to necessitate a short-range corrective term) we choose the traditional Buckingham form for the short-range corrective part interatomic potential [26]

$$V_{ij} = Ae^{\frac{-r_{ij}}{\rho}} - \frac{C}{r_{ij}^6} \quad (1)$$

in addition to an ion-ion coulombic term without cutoff. The first term of amplitude $A$ and range $\rho$ aims at describing the quantum exclusion principles preventing the interpenetration of the electronic clouds of the atoms, when the second term of amplitude $C$ comes from a classical dipole-dipole interaction supposed to act in between atoms. A fit of interatomic potentials on structural data is not sufficient to describe the specificities of ferric fluoride. On the basis of a recent estimation of the bulk modulus at 14 GPa [27] in the case of the rhombohedral ferric fluoride, the potential parameters have been fitted by means of the GULP program: the values are listed in Table 1. Those parameters allow to reproduce the lattice parameters and the cationic and anionic positions observed experimentally in bulk $MF_3$ (M= Fe, Ga, Co, Ru, Pd, and Ir) which crystallize in r-3c (167) space group, particularly those established at 300K on $FeF_3$ [28]. Besides, when this procedure is applied on a free iron fluoride molecule, we found a bond length of 1.72 Å, within experimental error [29], which reassures us on the structures obtained in or near a vacuum. The method presented in this paper in order to model GB allowed also to reproduce perfect rhombohedral ferric fluoride starting from two independent blocks without relative disorientation. It is important to emphasize that this result could not be attained using potentials such as those from [26] using a cutoff.

3. Computational Aspects

Our main goal is to describe the GB (intergranular) structures in ionic systems, thus extending a previous work devoted to a simpler system of bcc iron modeled by the Embedded Atom Method [30] where the GBs result from a dense random packing of hard spheres. In the present study, both the larger range of the potential and the maintenance of local stoichiometry make the problem computationally much more complex and time consuming.

3.1 Initial setup

We keep the same method as in our previous study for the initial setup of the simulation, which consists in choosing several fixed nucleation centers inside a simulation box. For each center, Euler angles are chosen in order to orient the relative crystallographic axes of the nanocrystallites. This arrangement aims at reproducing a polycrystalline structure modeling the one obtained by mechanosynthesis. The next step is, by adding atoms to the system, to make the crystallites grow in all directions until the structure reaches a certain limit (with a very large, arbitrary total number of atoms, as will be discussed further below). This limit is given by Voronoï cell conditions [31]: all the nodes belonging to a Voronoï cell are the closest from the nucleation center included in this cell compared to all other nucleation centers. We then use one of the two methods described in [30] to start with a system with a natural configuration. The first one allows the grains to overlap each other and to remove atoms closer to each other than a certain arbitrary threshold. On the contrary, the second approach prevents the grains from overlapping each other (separating them from the width of about one octahedral unit) and then bringing them in contact using a modified Metropolis procedure relaxing simultaneously the position of the nucleation centers and the atoms themselves. An *ad hoc* choice of both temperature and displacement step size allows the ejection of some atoms from the initial system in order to keep the stoichiometry as homogeneous as possible between atomic species and to maintain the architecture based on octahedral units. Besides, when the crystallites are brought together, only atoms close to the interface, (determined from the Voronoï conditions and the nucleation centers), have to be relaxed. This step is successfully achieved by submitting atoms to the annealing procedure with probability $\exp(-\alpha x)$, where x is the distance between the considered atom and the closest interface and $\alpha = \alpha_i + 3(D_i-D)/D_i^2$ where $D_i$ is the initial distance in between crystallites and D the distance corresponding to the current step of the simulation. The choice of $\alpha_i$ does meet two conditions: (i) when $D=D_i$, $\alpha$ must be large enough in order to relax atomic positions at the interface of each grain over a length scale of at least one octahedral unit, and (ii) when D goes to zero, the values of $\alpha$ should tend to a value comprised between 0.5 and 2 Å$^{-1}$, as used in the final annealing discussed below. A full disk storage of the energies and structures allows to pinpoint the lowest energy configuration, which can appear with or without interpenetration of the grains. The structure at this stage has reached a minimum but local defects might occur, especially at the interface. Those procedures are admittedly unphysical and arbitrary but lead to physically acceptable (i.e. stable) and non-discernable (i.e. using observables such as the pair distribution function) states after further annealing as described below.

3.2 Annealing schedule

We use the Monte Carlo simulated annealing scheme with the Metropolis algorithm [32]. The system is maintained at 200K until equilibrium is achieved, as measured by an equal ratio of accepted and rejected moves, and finally relaxed down to 0K until a local minimum is reached to be free from thermal effects. This typically takes 300 iterations of 100000 Monte Carlo steps done at constant temperature. As previously discussed in the introduction, this scheme actually works so well that it rapidly leads to the lowest free energy configuration for the system, namely a quasi-infinite piece of bulk if periodic boundary conditions are used. In addition, the size of the simulation box is chosen to reproduce zero or close to zero pressure, or a large cluster when such periodic conditions are not enforced [33]. One could however find certain commensurate tilt and twist angles of grains compatible with a certain range of simulation box sizes, but on the one hand such a search is mathematically very complicated in three dimensions [34] (not mentioning the problem of triple GB), and on the other hand we want to address arbitrarily oriented GBs. We therefore use the following modification to the Metropolis procedure, close to the one discussed for the setup of the system: the probability for a given atom to be subjected to the Metropolis algorithm is given by $P(x) = e^{-ax}$, where *x* is the distance from the considered atom to the nearest interface and *a* an adjustable parameter ranged from 0.5 to 2 Å$^{-1}$. This probability law is simulated by comparison to a random deviation drawn between 0 and 1 according to Von Neumann's rejection method. In the core of the GB where the probability of moving atoms is larger, the atomic structure is then much more disordered and the disorder decreases gradually while moving away from the interface tends toward a fixed, constrained structure at large distance from the interface.

The choice of P($x$) can be alternatively seen as an optimal way to enforce arbitrary constraints for the system considered. At infinite distance, crystallites are forced to be described as perfect atomic lattices; however with arbitrary relative orientations to which the system will tend to adapt, we therefore have a way to describe any GB in three dimensions with a quasi-infinite total number of atoms, without any periodic boundary conditions. Since P($x$) goes to zero far from the interfaces, at those distances atoms will almost never be subjected to a Monte Carlo shift: in this way we can have a quasi-infinite number of atoms in our simulation, smoothly enforcing boundary conditions. As can be seen in figure 1, using a cutoff of the coulombic potential and grains larger than this size would lead to unphysical oscillations in the total energy and an absence of convergence of the procedure since the grains would be brought together as more or less atoms would enter the computation contributing energies comparable to the changes resulting from the Monte Carlo moves. In the typical simulations presented below, we consider a full coulombic part of the potential without cutoff or approximation, providing thus a hint to the simplifying power of the method. The total energy due to all atoms is computed, but in the annealing procedure, due to the additive nature of the two-body contributions, only relative changes are computed when the position of one atom is changed in a trial move. It is important to remember that the surface effects in the total energy remain present: but since outer surface atoms will almost never move and stay in the bulk configuration, they do only contribute to a global shift of the total energy.

Besides, we have checked as in [30] that the choice of methods to establish the initial state has no significant effect on the structure occurring at the interface. Various reasonable values of parameter $a$ (range of the exponential) were tested in order to emphasize its small influence on the final result, as a does not exceed two inverse atomic lattice parameters. Two opposite situations have to be considered according to the value of a parameter: (i) when $a$ is too small (<0.5 Å), the interface turns out too sharp and therefore leads to an absence of relaxation of the outer core of the grains due to the interface; (ii) when $a$ is rather large (>2 Å) the simulation slows down and leads to a perfect cluster without a GB as all the atoms participate in the annealing scheme. The results can therefore be considered independent of either the value of $a$ or of the method used to construct the initial condition and constrain the system, if parameter $a$ is varied within a reasonable range, i.e. between 0.5 and 2 Å.

4. Results and discussion

The following results have been obtained after convergence of the Metropolis annealing procedure, in a measuring box of size half the total size of the system in order to avoid the effect of the edges. We present, as typical of our results, a system in a cubic box of 60 nm size, consisting of 16609 atoms distributed into two crystallites with Euler angles relative to the laboratory referential 30, 45 and 38 degrees for the left one and -28, 30 and -25 for the right one, showing that the present method allows thus to address grain boundaries in between crystallites with arbitrary relative orientations, or a bicrystal. Parameter $\alpha_i$ was fixed at 0.98 Å$^{-1}$. Systematic explorations of results as a function of the relative tilt and twist disorientations of the grains show energetic trends comparable to the ones of [34] but do not lead to qualitative differences in topological order except for trivial situations (perfect match or 45 degrees tilt or twist for instance).

4.1 Structural aspects and comparison to previous work

Since the crystalline grains are constrained to be perfect at a long distance from the interface thanks to the modified Metropolis scheme, the mismatch in orientation of the grains originates an excess of energy favoring the development of a structural change within the grain boundary. The morphology of the octahedral units is however preserved, as illustrated by the distribution of Fe-F distances and of superexchange Fe-F-Fe angles in figure 1. In addition one can also infer from Fig. 2 the size of the GB from the half-width of the variation of the Fe-F distance and the superexchange angle as a function of the distance to the interface. The thickness of the grain boundaries is estimated at about 1 nm, which is consistent with experimental structural data [5]. This confirms that the long range of the coulombic potential is necessary to reproduce a long-range order of the system and to preserve a corner sharing octahedral ordering. On the contrary, when the cutoff of the potential is too small (less than 2 nm), one only locally preserves $FeF_6$ octahedral units, but they tend either to form a perfect bulk if no topological disorder is introduced, or to turn into an amorphous piling with random disorder with respect to each other. This can be desired when describing a glassy state, but in this case no order of the iron atoms remains, in contradiction with magnetic measurements in nanostructured powders such as the ones of [5, 17]. Fig. 3 illustrates the radial distribution functions established from the GB, in comparison with those of crystalline grains. The present results are consistent with those resulting from either with simple electrostatic-like simulations such as [18] or those issuing from the simulation of random packing of perfect octahedral units [19,20]. It can be concluded that the same evolution of the superexchange angle is observed when the topological disorder is introduced, but the present method allows to describe the smooth transition in between perfectly ordered zones in crystallites and disordered grain boundaries. It is finally important to emphasize that the results of the present work, if limited to the GB, however confirms the validity of the past approaches for totally disordered systems.

4.2 Ring statistics analysis

Because the cationic topology behaves as a relevant probe of the disorder, i.e. the magnetic structure, we establish the ring statistics (defined as histograms of topological loops in between neighboring identical atoms closer to each other than a cut-off radius found from the first minimum of the pair distribution function), from nodes located within the GBs. The ring statistics is compared in Table 2 with those obtained from hand-built and from computed amorphous models which consist of dense random packing of corner-sharing octahedral units [18-20]. One observes in our present data a rather large proportion of 4-cycles because the GB is highly constrained to neighboring crystalline grains. The presence of odd-membered rings in the GB does give rise thus to partially frustrated systems with independent antiferromagnetic crystalline grains linked by speromagnetic-like GBs, in agreement with experimental data [5, 17]. The simulated magnetic structures computed from the structures found with the present methods will be detailed in a forthcoming publication together with the estimate of interface magnetic anisotropy.

In order to quantitatively analyze the consequences of the modeled structures on magnetic properties, we start by looking for topological defects among the iron atoms as compared to the perfect bulk. We were able to localize 3, 4, 5, 6 and 7 membered rings in the system. Figure 4 illustrates the position of such atoms in the final structural system. Indeed, magnetic properties of such systems can be strongly modified by the presence of odd iron-iron topological cycles associated to antiferromagnetic superexchange coupling [35]. The system becomes then magnetically frustrated. Several compounds exhibit such an effect but they are either crystallographically totally ordered or completely topologically disordered. In our case the frustrated cationic rings are localized in or around the GB so that the magnetic structure remains antiferromagnetic far from the GB; however the proportion in volume of such atoms has important consequences on the magnetic properties of the system, destroying long-range antiferromagnetic order. This is in agreement with experimental results [5, 17].

5. Summary and conclusion

We have successfully extended a new method to compute the structure of nanoscale grain boundaries and interfaces to ionic systems with non-collinear magnetic properties, originated from the frustrated cationic topology. The present approach is based on simple modifications of the Metropolis algorithm inspired from previous work in nuclear or cluster collisions, allowing to address double and triple GBs with arbitrary relative disorientations without recurring to periodic boundary conditions. The number of simulation parameters is reduced after comparing the results of two methods to establish the initial condition of the simulation, and we end up with a much simpler scheme than molecular dynamics, which requires several further assumptions and also a higher numerical cost. The low cost of our methods allows using a full coulombic potential without cutoff, making possible to obtain any intermediary state in between two crystalline bulks or a crystalline block and an amorphous glass-like system, preserving the topology of the system. These results are fairly consistent with experimental structural data and magnetic measurements in mechanically milled nanostructured powders as well as with bulk under pressure. It is also important to emphasize that such a computational approach and previous one applied to metallic systems can be thus combined and developed to model all types of interface involving metallic, ionic and semi-conducting layer architectures or layers deposited onto different substrates. Once the structure is well established at the interface, the physical properties can be thus computed using different approaches allowing then understanding the role of interfaces after comparing numeric predictions to experimental data. We have for instance successfully coupled a classical Heisenberg magnetic simulation to the present results in another paper [36]

The various C++ and Fortran 90 programs used in this study are available upon request to the authors. BF was supported by the Service of cooperation of the French Embassy to Cameroon. We acknowledge collaboration with EG@ (Euro-Graduation-Access). Some computations were performed at IDRIS under grant i2009096171.

[1] H. Gleiter, Acta. Mater. 48 (2000).
[2] C. Suryanarayana and C. C. Koch, Hyp. Interact. 130 (2000).
[3] H. Gleiter, Prog. Mater. Sci. 33, 223.
[4] R. W. Siegel, Nanostruct. Mater., 3, 1 (1993).
[5] H. Guérault and J. M. Greneche, J. Phys.: Condens. Matter. 12 4791 (2000)
[6] A. Hernando, M. Vasquez and D. Paramo, Mater. Sci. Forum 269-272 1033 (1998)
[7] S. R. Agnew and J. R. Weertman, Mater. Sci. Eng. A 242, 174 (1998)
[8] R. Z. Valiev, I. V. Alexandrov, W. A. Chiou, R. S. Mishra and A. K. Mukherjee, Mater. Sci. Forum 235-238, 497 (1997).
[9] C. C. Koch, D. G. Morris, K. Lu and A. Inoue, MRS Bull 24, 54 (1999).
[10] P. Keblinski, D. Wolf, S. R. Phillpot and H. Gleiter, Scr. Mater. 41, 631 (1999).
[11] A. Sawa, T. Fuji, M. Kawasaki, Y. Yokura, Appl. Phys. Lett. 88 232112 (2006).
[12] T. Fuji, M. Kawasaki, A. Sawa, Y. Kawazoe, H. Akoh, Y. Yokura, Phys. Rev. B 75 165101 (2007).
[13] H. Van Swygenhoven and A. Caro, Phys. Rev. B 58, 11246 (1998).
[14] O. A. Shenderova, D. W. Brenner and L. H. Yang, Phys. Rev. B 60, 7043 (1999).
[15] M. C. Payre, P.D. Bristowe and J.D. Joannopoulos Phys. Rev. Lett. 58 1348 (1987)
[16] F. Sansoz, J.F. Molinari, Act. Mater. 53, 1931-1944 (2005).
[17] B. Bureau, H. Guérault, G. Silly, J.Y. Buzaré and JM Greneche J. Phys.: Condens. Matter. 11 L423 (1999).
[18] J. M. D. Coey and P. J. K. Murphy, J. Non-Cryst. Solids 50 125 (1982)
[19] J. M. Greneche, J. Teillet and J. M. D. Coey, J. Non-Cryst. Solids 83 2 (1986)
[20] J. M. Greneche, J. Teillet and J. M. D. Coey, J. Physique 48 1709 (1987)
[21] J. D. Gale and A.L. Rohl, Mol. Simul., 29, 291 (2003)
[22] A. W. Appel, SIAM J.Sci. Stat. Comput. 6 85-103 (1985).
[23] J. Barnes, P. Hut, Nature 324, 446-449 (1986).
[24] L. Greengard, V. Rokhlin, J. Comp. Phys. 73 325-348 (1986).
[25] J. Board Jr., J. Causey, J. Leathrum Jr., Chem. Phys. Lett. 198, 89 (1992).
[26] C. Legein, J.Y. Buzaré, B. Boulard and C. Jacoboni, J. Phys.: Condens. Matter 7, 4829-4846 (1995).
[27] J.-E. Jorgensen and R. I Smith, Acta Cryst. (2006) B62, 987-992
[28] M. Hepworth, K. H. Jack, R.D. Peacock and G.J. Westland, Acta Cryst. (1957) 10, 63
[29] M. Hargittai, M. Kolonits, J. Tremmel, J.-L. Fourquet and G. Ferey, Struct. Chem. 1, 75 (1990)
[30] M. Grafouté, Y. Labaye, F. Calvayrac and J.M. Grenèche Euro. Phys. J. B 45 419 (2005)
[31] G. Z. Voronoi, J. Reine Angew, Math. 134, 199 (1908).
[32] F. Calvo and F. Spiegelman, Phys. Rev. Lett. 89, 266401 (2002).
[33] C. L. Cleveland and U. Landman, Science 257, 355-361 (1992).
[34] V. R. Coffman and J. P. Sethna Phys. Rev. B 77, 144111 (2008)
[35] J. M. Greneche and J. M. D. Coey, J. Physique 51, 231 (1990)
[36] B. Fongang, Y. Labaye, F. Calvayrac, S. Zekeng, J.M. Grenèche submitted to the Journal of Magnetism and Magnetic Materials 2010 (see also ArXiV)

**Figure 1**

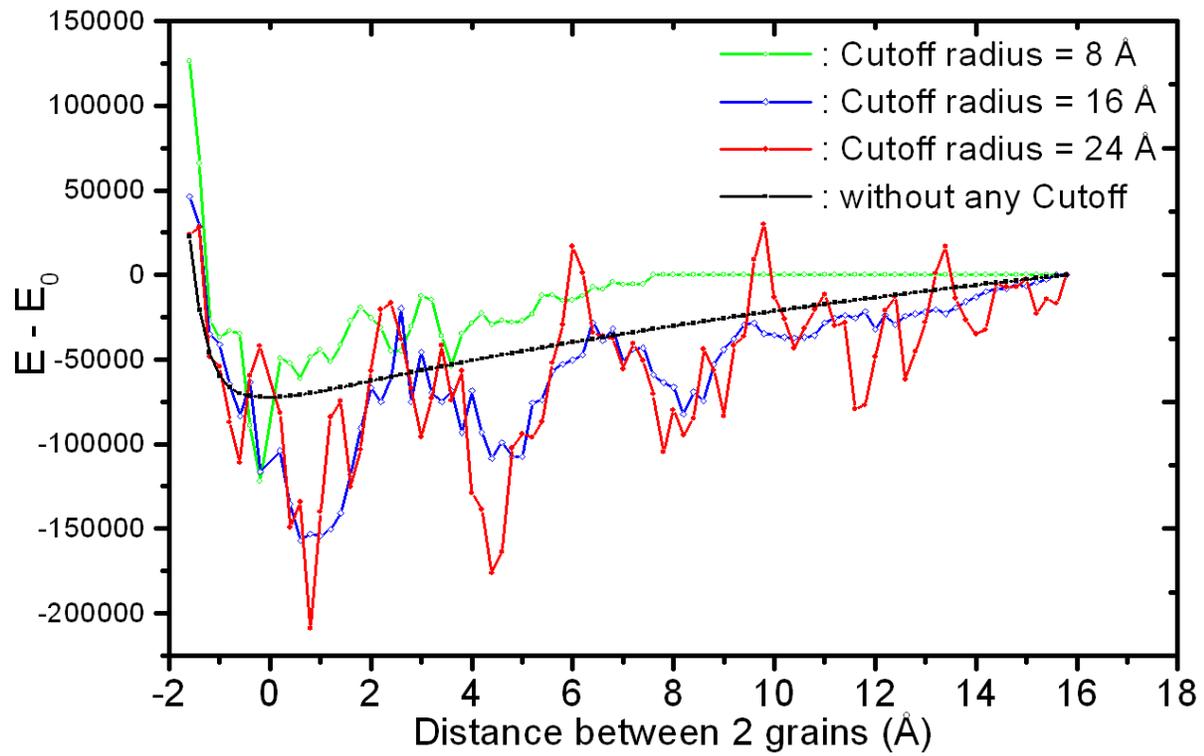

Figure 1 Evolution of binding energy of two grains as function of their mutual distance and cutoff radius of the interatomic potential .

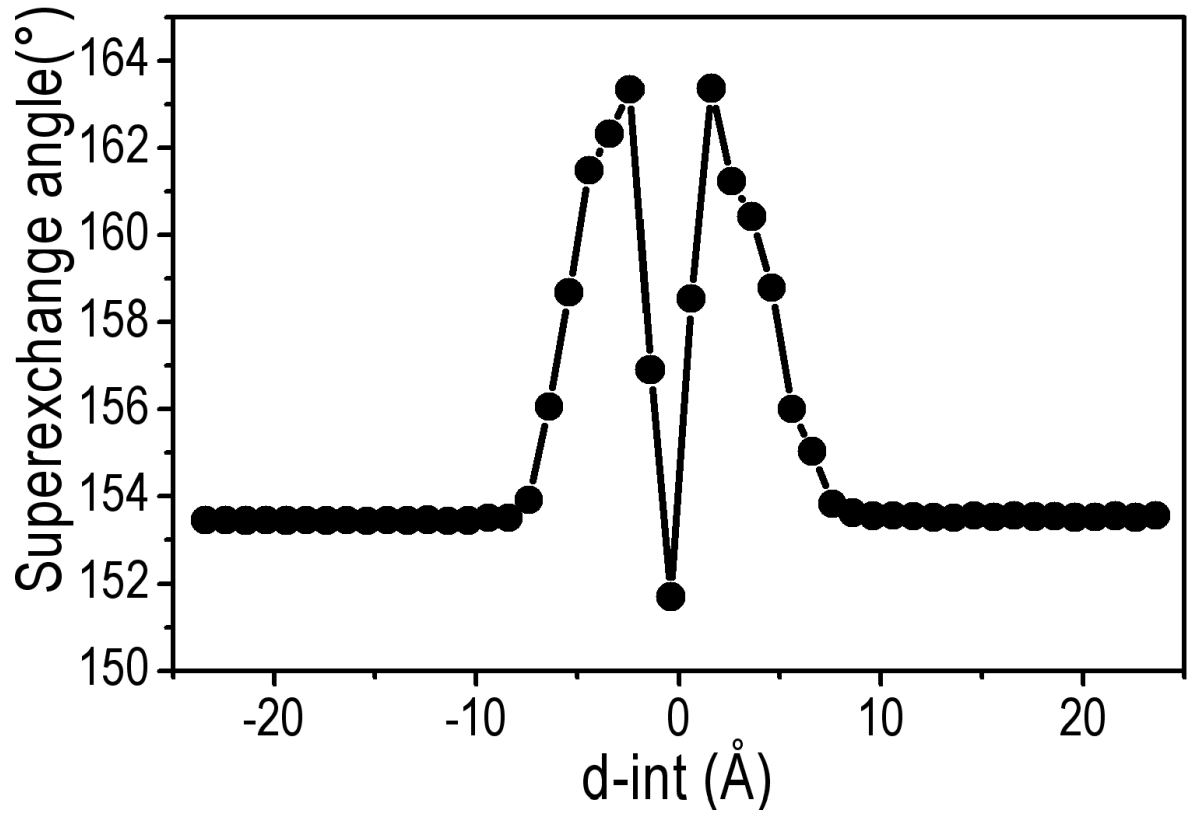

Figure 2 Distribution of superexchange angle as a function of the distance to the interface.

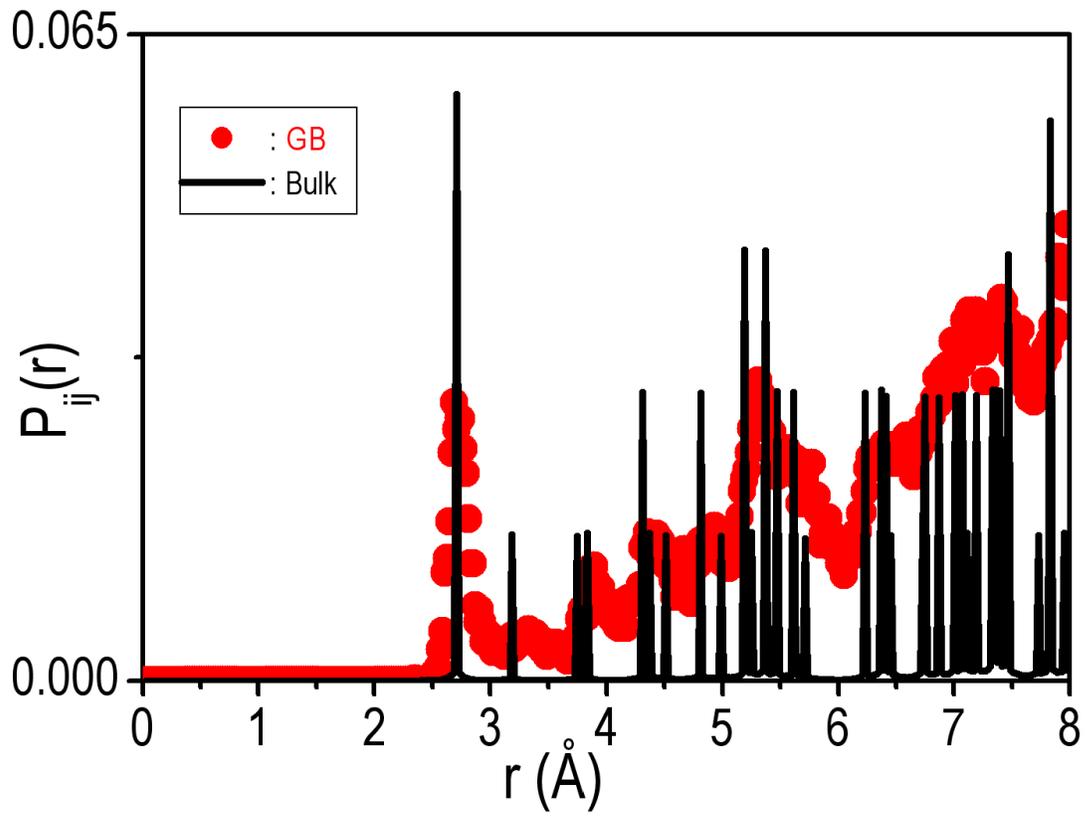

Figure 3 : Comparison of radial distribution functions for the F-F distances in the bulk and in the grain boundary

**Figure 4**

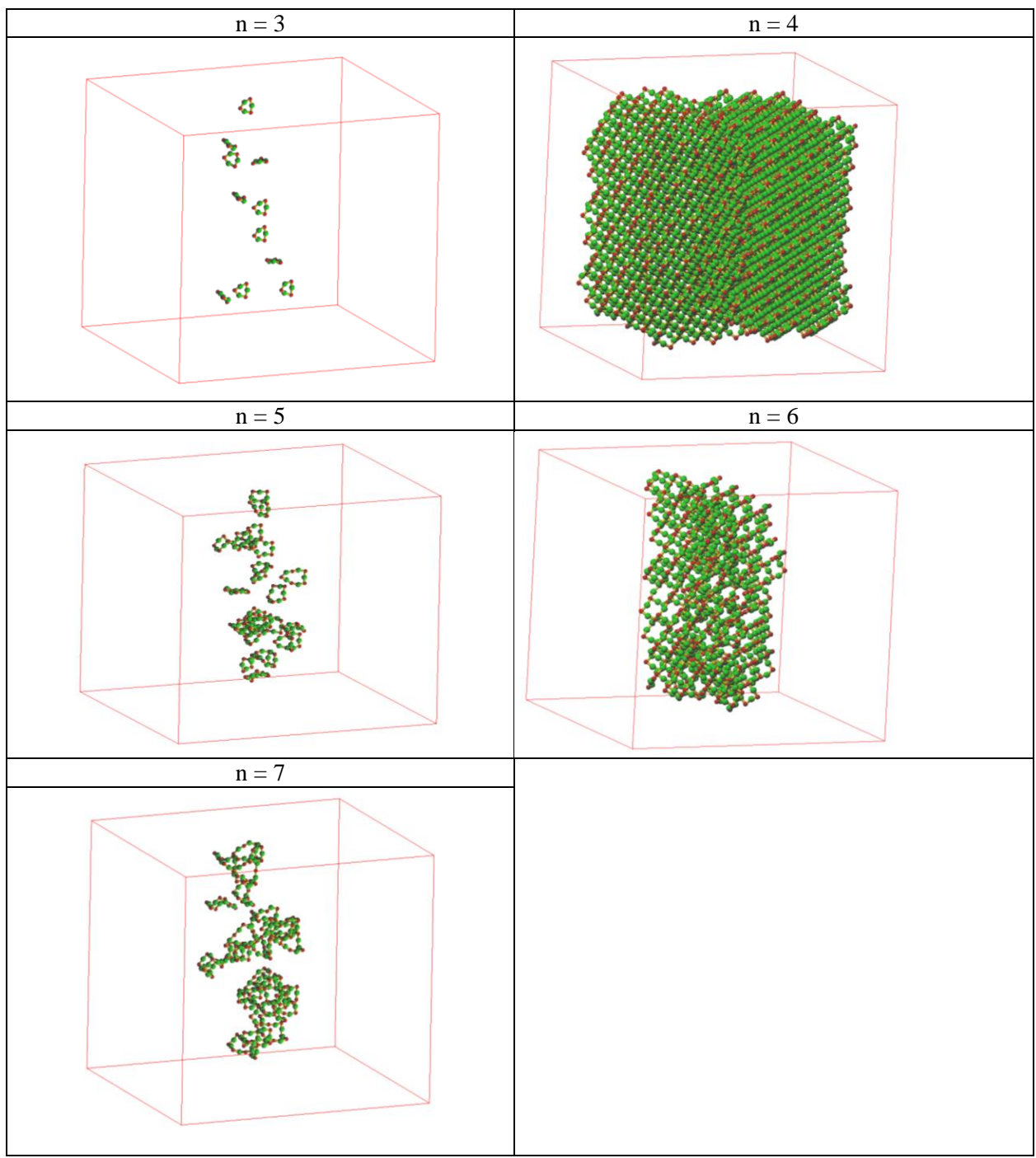

Figure 4: Distribution of n-membered rings of iron (n = 3, 4, 5, 6 and 7) in the simulation box. Odd-membered rings are only located at the interface.

**Table 1**

|       | A (eV)   | ρ $\rho$ (eV Å$^{-1}$) | C (eV Å$^{-6}$) |
|-------|----------|------------------------|-----------------|
| Fe - F | 2332.400 | 0.246                  | 0.000           |
| F - F  | 6242.496 | 0.274                  | 8.817           |

Table 1: Buckingham potential parameters

**Table 2**

| n-membered ring | n = 3 % | n = 4 % | n = 5 % | n = 6 % | n = 7 % | $<n>$ |
|---|---|---|---|---|---|---|
| Coey's model [18] | 8.1 | 46.5 | 42.2 | 2.5 | 0.7 | 4.41 |
| Octahedral units [19] | 15.9 | 38.7 | 37.9 | 7.5 | - | 4.38 |
| Computed model [20] | 28.0 | 36.3 | 31.2 | 3.6 | 0.9 | 4.14 |
| GB: present study | 9.5 | 55.7 | 12.3 | 15.3 | 7.2 | 4.55 |

Table 2: Ring statistics in GB compared to those of previous amorphous models (see text)